\documentclass[a4paper,11pt]{article}
\pdfoutput=1 

\usepackage{jheppub} 
\usepackage{enumitem}
\usepackage{mathrsfs}  

\usepackage{lscape}
\usepackage{booktabs}

\usepackage[T1]{fontenc} 
\usepackage{epsfig}
\usepackage{braket}
\usepackage[dvipsnames]{xcolor}
\usepackage{comment}
\usepackage{float}
\usepackage{caption}
\usepackage{subcaption}
\usepackage{bbold}  
\usepackage{hyperref}
\newcommand{\postscript}[2]{\setlength{\epsfxsize}{#2\hsize}
   \centerline{\epsfbox{#1}}}

\usepackage{amsmath}

\title{\boldmath New insights on a sign-switching $\Lambda$ }

\author[a]{Jorge F. Soriano,}  
\author[a,b]{Shimon Wohlberg,}
\author[a,b,c]{and Luis A. Anchordoqui}

\affiliation[a]{Department of Physics, Lehman College, City University of New York, NY 10468, USA}
\affiliation[b]{Department of Astrophysics, American Museum of Natural History, NY 10024, USA}
\affiliation[c]{Department of Physics, Graduate Center, City University of New York, NY 10016, USA}

\emailAdd{jorge.soriano@lehman.cuny.edu,
shimon.wohlberg@lc.cuny.edu, luis.anchordoqui@gmail.com}

\abstract{The proposal for a sudden sign-switching cosmological
  constant $\Lambda$ in the local universe, emulating a phase
  transition from anti-de Sitter (AdS) to de Sitter (dS) space, has
  markedly revamped the fit to observational data and lays out a
  propitious framework for ameliorating  major cosmological tensions,
  such as the $H_0$ and $S_8$ tensions. This proposal is widely known as $\Lambda_s$CDM. We investigate the possibility
  that $\Lambda$ does not only flip sign at the transition 
  but has also different curvature radii in the AdS and dS phases. We show that 
  the critical redshift of the transition $z_c$ is strongly correlated with the vacuum energy in the AdS phase $\Omega_{\Lambda_-}$, and that these two variables do not correlate strongly with the other cosmological parameters. We also show that the cost of adding an additional parameter to the $\Lambda_s$CDM cosmological model does not improve the goodness of fit. Armed with our findings, we demonstrate that for a proper choice of $z_c$,  the vacuum energy in the dS phase may not necessarily be $-\Omega_{\Lambda_-}$, for comparable degree of conformity between the model prediction and experimental data.}

\begin{document} 
\maketitle
\flushbottom

\section{Introduction}

$\Lambda$ cold dark matter ($\Lambda$CDM) has proven incredibly
successful at describing many features of cosmology that we observe in
our experiments~\cite{ParticleDataGroup:2024cfk}. The most famous example, arguably, is the agreement
with data from the cosmic microwave background (CMB)~\cite{Planck:2018vyg}.
Yet, with the increase in precision of cosmological observations over
the past decade, this success has been challenged. Two of the most
compelling shortcomings of $\Lambda$CDM are its predictions of the
present-day expansion rate $H_0$ and the amplitude of the matter clustering in
the late Universe (parameterized by
$S_8$)~\cite{Abdalla:2022yfr}. Strictly speaking, the values $H_0 =
67.4 \pm 0.5~{\rm km/s/Mpc}$
and $S_8= 0.834 \pm 0.016$ inferred from {\it Planck}'s CMB data assuming $\Lambda$CDM~\cite{Planck:2018vyg} are in
$\sim 5\sigma$ tension with $H_0 =73.04 \pm 1.04~{\rm km/s/Mpc}$ from the SH0ES distance ladder
measurement (using Cepheid-calibrated type-Ia
supernovae)~\cite{Riess:2021jrx,Murakami:2023xuy} and in $\sim 3\sigma$ tension with
$S_8= 0.766^{+0.020}_{-0.014}$ from the cosmic shear data of the
Kilo-Degree Survey (KiDS-1000)~\cite{Heymans:2020gsg},
respectively. Systematic effects do not seem to be responsible for these  discrepancies (see e.g.~\cite{Scolnic:2024hbh}), which have
become a new cornerstone of theoretical physics. Naturally, many new physics 
setups are rising to the challenge~\cite{DiValentino:2021izs,Schoneberg:2021qvd,Perivolaropoulos:2021jda}.

$\Lambda_s$CDM~\cite{Akarsu:2019hmw, Akarsu:2021fol,Akarsu:2022typ,Akarsu:2023mfb} is one of the many new physics setups that have
been proposed to simultaneously accommodate the $H_0$ and $S_8$
tensions.\footnote{An alternative model that simultaneously addresses $H_0$ and $S_8$ is discussed in~\cite{Gomez-Valent:2024tdb,Gomez-Valent:2024ejh}.} The setup relies on an empirical conjecture which postulates
that $\Lambda$ may have switched sign (from negative to positive) at
critical redshift $z_c \sim 2$;
 \begin{equation} \Lambda\quad\rightarrow\quad\Lambda_{\rm s}\equiv \Lambda_0 \ {\rm sgn}(z_c-z),
\label{Lambdas}
 \end{equation}
 with $\Lambda_0>0$, and where ${\rm sgn}(x)=-1,0,1$ for $x<0$, $x=0$
 and $x>0$, respectively. Apart from resolving the three major cosmological
 tensions, $\Lambda_s$CDM achieves quite a good fit to Lyman-$\alpha$
 data provided $z_c \lesssim 2.3$~\cite{Akarsu:2019hmw}, and it is in agreement with the otherwise puzzling JWST
observations~\cite{Adil:2023ara,Menci:2024rbq}.

The colossal success of
$\Lambda_s$CDM in accommodating observations of baryon acoustic
oscillations (BAO)  is in effect contingent on the (angular)
transversal two dimensional (2D) BAO data on the shell, which are less
model dependent than the 3D BAO data. This is because the 3D BAO data
sample relies on $\Lambda$CDM to determine the distance to the
spherical shell, and hence could potentially introduce a bias when
analyzing cosmological scenarios beyond
$\Lambda$CDM~\cite{Bernui:2023byc, Gomez-Valent:2023uof}. Indeed, by
using 2D BAO data the actual SH0ES $H_0$ measurement and the angular
diameter distance to the last scattering surface can be simultaneously
accommodated, but at the expense of having an effective negative energy
density  for $z \gtrsim 2$.

Another caveat associated to $\Lambda_s$CDM is that the instantaneous
transition of the vacuum energy density, which is governed by the
signum function, gives rise to a hidden sudden singularity at
$z_c$~\cite{Barrow:2004xh}. Note that the scale factor $a$ of
$\Lambda_s$CDM is continuous and non-zero at the critical time $t = t_c$, but its first derivative $\dot a$ is discontinuous, and its second derivative $\ddot a$ diverges. However, the sudden singularity yields a minimal impact on the formation and evolution of cosmic bound structures, thereby preserving the viability of $\Lambda_s$CDM~\cite{Paraskevas:2024ytz}.

The changeover of $\Lambda$ at $z_c$ can be
interpreted as a phase transition
between anti-de Sitter (AdS) and (dS) spaces, because $n$-dimensional dS and AdS spaces can be
regarded as solutions of Einstein's field equations
of empty spaces with cosmological constant
\begin{equation}
  \Lambda = (n-2)R_i/(2n) \,,
\end{equation}  
where $i= \{{\rm
    dS}_n, {\rm AdS}_n\}$, $R_{{\rm dS}_n} = n (n-1)/\ell_{{\rm dS}_n}^2$ and $R_{{\rm AdS}_n} = - n
(n-1)/\ell_{{\rm AdS}_n}^2$ are respectively the dS and AdS Ricci scalars, and
$\ell_i$ is the radius of curvature~\cite{Hawking:1973uf}.

The essence of the instantaneous sign-switching cosmological constant
posits a challenging puzzle in identifying a concrete theoretical
model able to accommodate the AdS $\to$ dS transition. The puzzle is actually exacerbated by the AdS distance conjecture, which states that 
there is an arbitrarily large distance between AdS and dS vacua in metric 
space~\cite{Lust:2019zwm}. Having said that, the phenomenological success of $\Lambda_s$CDM,
despite its simplistic structure, provides robust motivation to search
for possible underlying physical mechanisms to describe the
conjectured AdS $\to$ dS transition. Many theoretical realizations
have been
proposed~\cite{Anchordoqui:2023woo,Akarsu:2024qsi,Anchordoqui:2024gfa,Akarsu:2024eoo,Yadav:2024duq,Toda:2024ncp,Dwivedi:2024okk,Anchordoqui:2024dqc,Akarsu:2024nas,Souza:2024qwd}. Motivated
by the phenomenological success of $\Lambda_s$CDM, in this paper we investigate the possibility
that $\Lambda$ does not only flip sign at the transition but has also
different curvature radii in the AdS and
 dS phases. In plain English, the AdS radius $\ell_{{\rm AdS}_4}$
 characterized by $\Lambda_-$ and the dS radius $\ell_{{\rm dS}_4}$
 characterized by $\Lambda_+$ satisfy $|\Lambda_-| \neq
 |\Lambda_+|$.

 The layout of the paper is as follows. In Sec.~\ref{sec:2} we introduce the cosmological parameters explored in this work. In Sec.~\ref{sec:3} we describe the analysis method and the observational
datasets/likelihoods used to constrain the cosmological
parameters. After that 
we implement the numerical computations, present our results, and
discuss their implications. The
paper wraps up in Sec.~\ref{sec:4} with some conclusions.

\section{Cosmological Parameters} 
\label{sec:2}

There are two basic hypotheses in modern cosmology: {\it (i)}~the assumption of the validity of General Relativity and {\it (ii)}~the cosmological principle, which is the assumption that the universe on cosmological scales is homogeneous an isotropic. Together, these two hypotheses set constraints on the four-dimensional spacetime metric, which reduces to the maximally-symmetric Friedmann-Lama\^{\i}tre-Robertson-Walker line element
\begin{equation}
  ds^2 = -dt^2 + a^2(t) \left[ \frac{dr^2}{1-kr^2} + r^2 (d\theta^2 + \sin^2 \phi \ d\phi^2) \right] \,,
\label{metric}
\end{equation}
where $(t, r, \theta, \phi)$ are co-moving coordinates,  $k (= -1, 0, 1)$ parametrizes the curvature of the homogeneous and isotropic spatial sections, and $a(t)$ is the cosmic scale factor, which is related to the redshift $z$ by $a = 1/(1+z)$~\cite{Weinberg:2008zzc}. Observations favor a spatially flat $(k=0)$ universe~\cite{ParticleDataGroup:2024cfk}. The expansion rate of the Universe is measured by the Hubble parameter,
\begin{equation}
  H = \frac{\dot {a}}{a} \, ,
\label{HubbleP}
\end{equation}  
and its present-day value is known as the Hubble constant
$H_0 = 100~h~{\rm km/Mpc/s}$, with $0< h < 1$~\cite{Hubble:1929ig}. 

The widely accepted spatially-flat $\Lambda$CDM model requires only 6
independent parameters 
\begin{equation}
{\cal P}_{\Lambda{\rm CDM}}  = \{\omega_b,\, \omega_{cdm},\,
\theta_s, \, \tau_{\rm reio}, \,  n_s, \, A_s\} \,,
\label{P1}
\end{equation}
to completely specify the cosmological evolution, where $\omega_b
\equiv \Omega_b h^2$ is the baryon density, $\omega_{cdm} 
\equiv \Omega_c h^2$ is the CDM density, $\theta_s$ is the angular size of the sound
horizon at recombination, $\tau_{\rm reio}$ is the Thomson scattering
optical depth due to reionization, $n_s$ is the scalar
spectral index, and $A_s$ is the power spectrum amplitude of adiabatic
scalar perturbations. The $\Omega_i$ parameters are defined as the
ratio of the present day mean density of each component $i$ to the
critical density; by definition $\sum_i \Omega_i = 1$. Neglecting smaller order terms, $\Omega_c + \Omega_b + \Omega_\Lambda \sim 1$. 

The specific set of six parameters used to define the cosmological
model is somewhat open to choice. Within the context of fitting a
$\Lambda$CDM model to a CMB power spectrum, the six selected key
parameters are primarily chosen to avoid degeneracies and thus speed
convergence of the model fit to the data~\cite{Kamionkowski:1999qc,Hu:2001bc,Kosowsky:2002zt}. Other
interesting parameters providing additional physical insight may be
derived from the model once the defining six parameters have been
set. These include:
\begin{itemize}[noitemsep,topsep=0pt]
\item The present-day expansion rate $H_0$.
\item The
amplitude of the matter clustering in the late universe, parameterized
by 
\begin{equation}
S_8 \equiv \sigma_8
\left(\frac{\Omega_m}{0.3}\right)^{0.5} \,,
\end{equation}
where $\sigma_8$ is the root mean square of the
amplitude of matter perturbations smoothed over $8h^{-1} {\rm Mpc}$
and $\Omega_m = \Omega_c + \Omega_b$ is the non-relativistic matter density.
\item The effective number of relativistic neutrino-like species 
  $N_{\rm eff}$ is a convenient parametrization of the relativistic
  energy density of the universe beyond that of photons, in units
  of the density of a single Weyl neutrino species~\cite{Steigman:1977kc}. Using conservation of entropy, fully thermalized relics with $g_*$ degrees of freedom contribute
\begin{equation}
  N_{\rm eff} = 3.043  + g_* \left(\frac{43}{ 4g_s}\right)^{4/3} \left
    \{ \begin{array}{ll} 4/7 & {\rm for  \ bosons}\\ 1/2 & {\rm for \
                                                           fermions} \end{array}
               \right.                                        \,,
\end{equation}    
where 3.043 is the standard-model benchmark~\cite{Cielo:2023bqp} and $g_s$ denotes the effective degrees of freedom for the entropy
of the other thermalized relativistic species that are present when
they decouple~\cite{Anchordoqui:2011nh}.\footnote{We adopt the most
  recent NLO calculation of $N_{\rm eff}$ in the standard model~\cite{Cielo:2023bqp}, which
 yields a reduction from  previous estimates~\cite{Mangano:2005cc,deSalas:2016ztq,Bennett:2019ewm,Akita:2020szl,Froustey:2020mcq,Bennett:2020zkv}.}
\end{itemize}

Herein, we aim to explore possible deviations from  $\Lambda$CDM. To
quantitatively assess these deviations we study the $\Lambda_s$CDM 7-parameter model, 
\begin{equation}
{\cal P}_{\Lambda_s{\rm CDM}} = \{\omega_b,\, \omega_{cdm},\,
\theta_s, \, \tau_{\rm reio}, \,  n_s, \, A_s, \, z_c\} \,,
\label{P2}
\end{equation}
and the $\Lambda_\mp$CDM 8-parameter model
\begin{equation}
{\cal P}_{\Lambda_\mp{\rm CDM}} = \{\omega_b,\, \omega_{cdm},\,
\theta_s, \, \tau_{\rm reio}, \,  n_s, \, A_s, \, z_c, \, \Omega_{\Lambda_-}\}
\, .
\label{P3}
\end{equation}
As might be expected, for both these models the first six parameters
are common ones with $\Lambda$CDM. 

Before proceeding, we pause to mention additional, non-cosmological parameters that are included in the model. First, there are 53 nuisance parameters needed to characterize the Planck and KiDS-1000 likelihoods, described in the next section. More importantly, the absolute magnitude of type Ia supernovae $M_B$ (needed for the Pantheon+ likelihood; see next section), must be considered too. It has been suggested that the $H_0$ tension is actually a tension on $M_B$~\cite{Camarena:2021jlr,Efstathiou:2021ocp}, because the base $\Lambda$CDM cosmology inferred from {\it Planck}  leads to $M_B=-19.401 \pm 0.027~{\rm mag}$~\cite{Camarena:2019rmj}, whereas the the SH0ES Cepheid photometry and Pantheon supernova peak magnitudes yield $M_B =-19.244 \pm 0.037~{\rm  mag}$~\cite{Efstathiou:2021ocp}. It has been shown that $\Lambda_s$CDM cosmology provides a robust solution to the $M_B$ tension~\cite{Akarsu:2023mfb} .

We consider two scenarios with
$N_{\rm eff} = 3.043$ and $N_{\rm eff} = 3.294$. The latter characterizes a
model that explains the AdS $\to$ dS transition using the contribution
to the Casimir energy of fields (the graviton, a real scalar, and three
right-handed neutrinos) propagating in the bulk of a five-dimensional spacetime~\cite{Anchordoqui:2023woo}.  The scalar field
has a potential holding two local minima with very small difference in
vacuum energy and bigger curvature (mass) of the lower one, and thus
when the false vacuum tunnels to its true vacuum state, the field
becomes more massive and its contribution to the Casimir energy
becomes exponentially suppressed~\cite{Anchordoqui:2024dqc}. The tunneling process then changes
the difference between the total number of fermionic and bosonic
degrees of freedom contributing to the quantum corrections of the
vacuum energy in a way that leads to  a sign-switching cosmological
constant at $z_c \sim 2$.

\section{Statistical Methodology,  Observational Data, and Numerical Analysis}
\label{sec:3}

We adopt the Markov Chain Monte Carlo (MCMC) technique. A
(discrete-time)
Markov chain with (finite or countable) state space  $\mathscr{X}$ consists of a
sequence $X_0, X_1, \cdots$ of $\mathscr{X}$-valued random variables such that for all states $i,j,k_0,k_1
\cdots$ and all times $n=0,1,2,\cdots$, 
\begin{equation}
  P(X_{n+1} = j \parallel X_n=i, X_{n-1} = k_{n-1}, \cdots) = p(i,j)
\end{equation}
where $p(i,j)$ depends only on the states $i,j$, and not on the time
$n$ or the previous states $k_{n-1},k_{n-2},\cdots$. The numbers $p(i,j)$
are called the transition probabilities of the chain. Note that each number is stochastically obtained from the previous number without explicitly being dependent on it. In our
analysis the states of a chain consist on specific values of the free
parameters in a given cosmological model.

To run the MCMC we adopt the {\tt MontePython} code, which interfaces
with the Cosmic Linear Anisotropy Solving System {\tt
  CLASS}~\cite{Blas:2011rf,Lesgourgues:2011re, Brinckmann:2018cvx}.
To explore the full parameter space of the models characterized by
(\ref{P2}) and (\ref{P3}) we have modified {\tt CLASS} to introduce
the possibility of an AdS $\to$ dS transition. We derive constraints
on the cosmological parameters of these two models and $\Lambda$CDM
from data-sets and likelihoods given below using the
Metropolis-Hastings algorithm~\cite{Robert:2015}, while enforcing the so-called 
 Gelman-Rubin convergence criterion of $R - 1 < 10^{-2}$ in all the
runs~\cite{Gelman:1992zz}. We make use of a series of astrophysical
and cosmological probes to construct our reference dataset:
\begin{itemize}[noitemsep,topsep=0pt]
\item \textbf{2018 Planck CMB data}: The CMB temperature,
  polarization, and lensing angular power spectra from the {\it Planck} 2018 legacy release~\cite{Planck:2018vyg,Planck:2019nip}. 
\item \textbf{Transversal BAO}: Measurements of 2D baryon acoustic
  oscillations (BAO),
 $\theta_{\text{BAO}}(z)$, obtained in a weakly model-dependent
 approach, and compiled in Table~I of~\cite{Nunes:2020hzy}.
\item \textbf{Pantheon+}: The 1701 light curves of 1550 distinct
  supernovae type Ia, which are distributed in the redshift interval
  \mbox{$0.001 \leq z \leq 2.26$}~\cite{Brout:2022vxf}. We incorporate
  the most recent SH0ES Cepheid host distance anchors~\cite{Riess:2021jrx} into the
  likelihood function by integrating distance modulus measurements of
  the Pantheon+ supernovae.
\item \textbf{Cosmic Shear}: KiDS-1000
  data~\cite{Kuijken:2019gsa,Giblin:2020quj}, including the weak
  lensing two-point statistics data for both the auto and
  cross-correlations across five tomographic redshift
  bins~\cite{Hildebrandt:2020rno}. We follow the KiDS team analysis
  and adopt the COSEBIs (Complete Orthogonal Sets of E/B-Integrals)
  likelihood~\cite{KiDS:2020suj}. Non-linearities are implemented into the
analysis using {\tt
  HALOFIT}~\cite{Smith:2002dz,Bird:2011rb}.\footnote{This
  implementation comes up with a
  difference to the data analysis carried out in~\cite{Akarsu:2023mfb}
  (which was executed using the  
 {\tt HMcode}~\cite{Mead:2015yca}), but such a difference leads to
 (almost) negligible effects in the results.}
\end{itemize}

\begin{figure}[tbh]
  \postscript{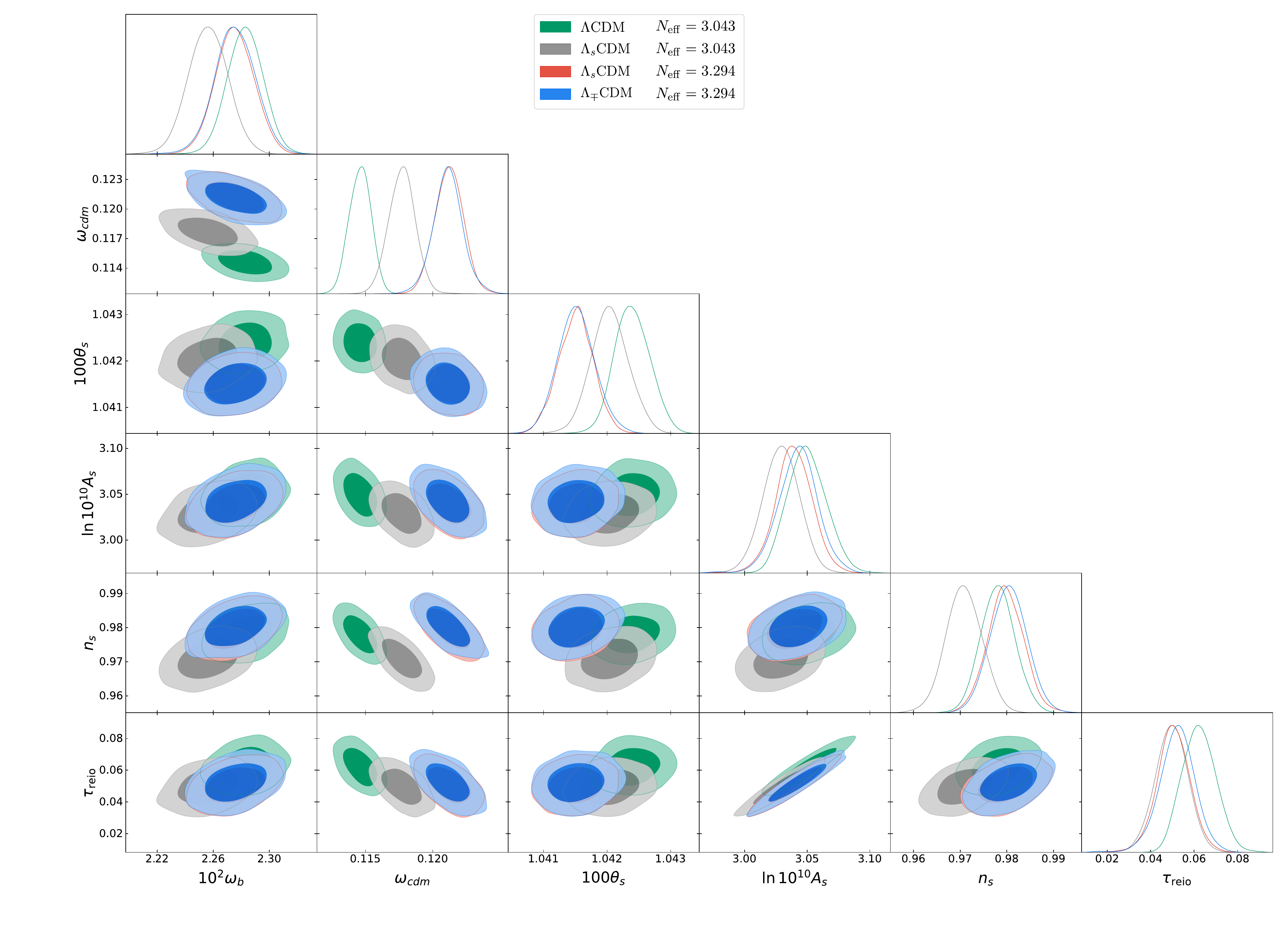}{0.99}
  \caption{Triangle plot showing two-dimensional contours at 68\% and 95\% CL and one-dimensional posterior probability distribution functions of the six parameters which are common to $\Lambda$CDM, $\Lambda_s$CDM, and $\Lambda_\mp$CDM. \label{fig:1}}
\end{figure}

\begin{figure}[tbh]
  \postscript{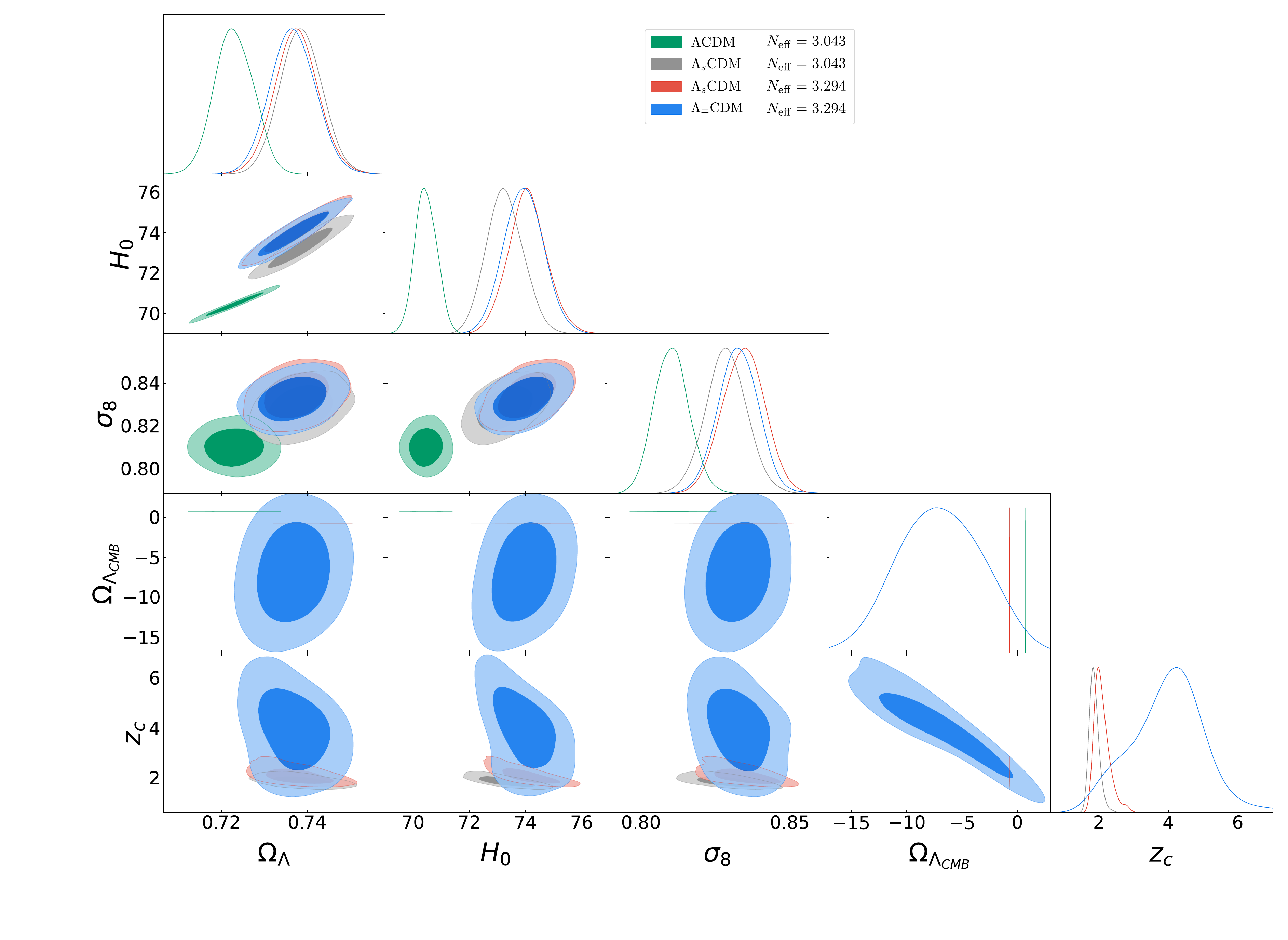}{0.92}
  \caption{Triangle plot showing two-dimensional contours at 68\% and 95\% CL and one-dimensional posterior probability distribution functions of $H_0$, $\sigma_8$, $\Omega_\Lambda$, and $\Omega_{\Lambda_{\rm CMB}}$ for $\Lambda$CDM, $\Lambda_s$CDM, and $\Lambda_\mp$CDM. \label{fig:2}}
\end{figure} 

Our results are encapsulated in Figs.~\ref{fig:1} and~\ref{fig:2}, and Table~\ref{tabla1}. There, for the sake of compactness, $\Omega_\Lambda$ has its standard meaning for $\Lambda$CDM and $\Lambda_s$CDM, and it is $\Omega_{\Lambda_+}$ for $\Lambda_\mp$CDM. Similarly, $\Omega_{\Lambda_{\rm CMB}}$ stands for $\Omega_\Lambda$ in $\Lambda$CDM, for $-\Omega_\Lambda$ in $\Lambda_s$CDM, and for $\Omega_{\Lambda_-}$ in $\Lambda_\mp$CDM. For this reason,some of these parameters are not independent from each other (that is, they are $100\%$ correlated), making some of the numbers in Table~\ref{tabla1} redundant.

It is straightforward to see that the fitted parameters of $\Lambda_s$CDM and $\Lambda_\mp$CDM are all consistent with each other at 95\%CL, independent of the value of $N_{\rm eff}$. Moreover, for $N_{\rm eff}=3.294$, they are effectively indistinguishable in regards to the six base parameters, as well as the derived $H_0$ and $\sigma_8$.  This implies that scenarios with $N_{\rm eff} = 3.294$ can simultaneously resolved the $H_0$ and $S_8$ tensions. For $\Lambda$CDM, we find $\chi^2_{\rm min} = 4224.7$. For $\Lambda_s$CDM, the goodness of fit always improves with respect to $\Lambda$CDM; namely, if $N_{\rm eff} = 3.043$ we find $\chi^2_{\rm min} = 4192.08$, whereas if $N_{\rm eff} = 3.294$ we find $\chi^2_{\rm min} = 4195.94$.\footnote{There is a difference of decimal digits with respect to the $\chi_{\rm min}$ reported in~\cite{Anchordoqui:2024gfa}, where a LO $N_{\rm eff} = 3.044$ was used.} 
On the other hand, while there is an improvement in the chi-squared fit of  $\Lambda_\mp$CDM when compared to $\Lambda$CDM, there is no improvement in the $\chi^2_{\rm min}$ when $\Lambda_\mp$CDM with $N_{\rm eff} = 3.294$ is compared to $\Lambda_s$CDM.\footnote{We note in passing that for  $\Lambda_\mp$CDM cosmology with $N_{\rm eff} = 3.043$ we obtain very similar results, but with $\chi_{\rm min}^2 = 4191.92$.}  We conclude that by adding the curvature radius in the AdS phase as a free model parameter we do not make better the fit to the data. 

In Fig.~\ref{fig:2}, due to the large plot range needed to show the $\Omega_{\Lambda_\mathrm{CMB}}$ distribution for $\Lambda_\mp$CDM, the corresponding distributions for $\Lambda$CDM and $\Lambda_s$CDM apparently collapse into a line. Specifically, the $\Lambda$CDM (green) and $\Lambda_s$CDM (overlapping red and gray, not always distinguishable) \emph{lines} are approximately at 0.72 and -0.74.

It is easily seen that $\Omega_{\Lambda_-}$ and $z_c$ are highly correlated with each other, and show almost no correlation with the rest of the parameters. This means that changes in $\Omega_{\Lambda_-}$ can be made compatible with observational data by a judicious choice of $z_c$: an earlier AdS to dS transition allows for larger (in magnitude) $\Omega_{\Lambda_-}$ values. One can roughly quantify this relation by choosing a sample of the MCMC points with largest likelihood and modeling the relation between these parameters. For instance, choosing the top $1\%$ points by likelihood, this relation is captured by 
\begin{equation}
\Omega_{\Lambda_-} \approx  - \frac{1}{2} \ z_c^2 + 1.33 \, .
\end{equation}

We end with two observations:
\begin{itemize}[noitemsep,topsep=0pt]
\item From Table~\ref{tabla1} it is straightforward to see that both $\Lambda_s$CDM and $\Lambda_\mp$CDM provide a solution of the $M_B$ tension, for the two fiducial values of $N_{\rm eff}$.
\item Our results are consistent with and complement those recently reported in~\cite{Akarsu:2025gwi}.    
\end{itemize}

\section{Conclusions}

\label{sec:4}

$\Lambda_s$CDM extends the $\Lambda$CDM concordance model of cosmology by promoting the assumption of a positive cosmological constant $\Lambda$ to a rapidly sign-switching cosmological constant $\Lambda_s$, at critical redshift $z_{\rm c}$. The innovative   $\Lambda_s$CDM cosmology provides a profitable arena to simultaneously accommodate the $H_0$ and $S_8$ tensions. In this paper we
explored the possibility of a similar transition, but between AdS and dS phases with different curvature radii. We have shown that $z_c$ is strongly correlated with the vacuum energy $\Omega_{\Lambda_-}$ in the AdS phase, and that there is no linear association between these variables and the other cosmological parameters. We have also shown that the cost of adding an additional parameter to the $\Lambda_s$CDM cosmological model does not improve the goodness of fit. Armed with our findings, we have demonstrated that for a proper choice of $z_c$,  the vacuum energy in the dS phase may not necessarily be $-\Omega_{\Lambda_-}$, for comparable degree of conformity between the model prediction and experimental data.

\begin{landscape}
\begin{table}
\caption{Marginalized constraints, mean values with 68\%~CL (95\%~CL), on the free and some derived parameters of the $\Lambda$CDM, $\Lambda_s$CDM and $\Lambda_\mp$CDM  models for different values of ($N_{\rm eff}$). \label{tabla1}}
	\begin{tabular}{c c c c c}
\toprule
Model	&	$\Lambda\mathrm{CDM}(3.043)$ & $\Lambda_s\mathrm{CDM}(3.043)$ & $\Lambda_s\mathrm{CDM}(3.294)$ & $\Lambda_\mp\mathrm{CDM}(3.294)$\\
\midrule
$10^{2}\omega_\mathrm{b}$ &$2.283\pm_{0.013}^{0.013}\,(\pm_{0.025}^{0.024})$                &$2.256\pm_{0.014}^{0.014}\,(\pm_{0.028}^{0.028})$                &$2.275\pm_{0.014}^{0.014}\,(\pm_{0.027}^{0.027})$                &$2.276\pm_{0.014}^{0.014}\,(\pm_{0.029}^{0.028})$                \\
$\omega_\mathrm{cdm}$                                            &$0.1146\pm_{0.00083}^{0.0008}\,(\pm_{0.00154}^{0.00151})$        &$0.11771\pm_{0.00099}^{0.00093}\,(\pm_{0.00197}^{0.0019})$       &$0.1212\pm_{0.0011}^{0.001}\,(\pm_{0.002}^{0.002})$              &$0.1211\pm_{0.0011}^{0.001}\,(\pm_{0.0022}^{0.0023})$            \\
$100\theta_s$                                                    &$1.0424\pm_{0.0003}^{0.00026}\,(\pm_{0.00052}^{0.00055})$        &$1.04204\pm_{0.00029}^{0.0003}\,(\pm_{0.00056}^{0.0006})$        &$1.04151\pm_{0.00029}^{0.00027}\,(\pm_{0.00056}^{0.00054})$      &$1.04152\pm_{0.00029}^{0.00029}\,(\pm_{0.00056}^{0.0006})$       \\
$\ln{10^{10}A_s}$                                                &$3.05\pm_{0.016}^{0.015}\,(\pm_{0.029}^{0.03})$                  &$3.029\pm_{0.014}^{0.015}\,(\pm_{0.03}^{0.028})$                 &$3.039\pm_{0.014}^{0.015}\,(\pm_{0.03}^{0.029})$                 &$3.042\pm_{0.015}^{0.016}\,(\pm_{0.033}^{0.032})$                \\
$n_s$                                                            &$0.9781\pm_{0.0037}^{0.0035}\,(\pm_{0.007}^{0.0074})$            &$0.9708\pm_{0.0039}^{0.0039}\,(\pm_{0.0076}^{0.0076})$           &$0.9798\pm_{0.0037}^{0.004}\,(\pm_{0.0076}^{0.0078})$            &$0.9804\pm_{0.004}^{0.0039}\,(\pm_{0.0077}^{0.0077})$            \\
$\tau_\mathrm{reio}$                                             &$0.0624\pm_{0.0082}^{0.0077}\,(\pm_{0.0149}^{0.0151})$           &$0.0496\pm_{0.0073}^{0.0074}\,(\pm_{0.0155}^{0.0144})$           &$0.0505\pm_{0.007}^{0.0075}\,(\pm_{0.0158}^{0.0152})$            &$0.0521\pm_{0.0075}^{0.0083}\,(\pm_{0.0177}^{0.0159})$           \\
$\Omega_{\Lambda}$					                                &$0.723\pm_{0.0045}^{0.0047}\,(\pm_{0.0085}^{0.0085})$			     &$0.7386\pm_{0.0048}^{0.0049}\,(\pm_{0.0095}^{0.0096})$           &$0.7375\pm_{0.0049}^{0.0051}\,(\pm_{0.01}^{0.0103})$             &$0.7367\pm_{0.0051}^{0.0053}\,(\pm_{0.0101}^{0.0102})$\\
$\Omega_{\Lambda_\mathrm{CMB}}$                                  &$0.723\pm_{0.0045}^{0.0047}\,(\pm_{0.0085}^{0.0085})$            &$-0.7386\pm_{0.0049}^{0.0048}\,(\pm_{0.0096}^{0.0095})$          &$-0.7375\pm_{0.0051}^{0.0049}\,(\pm_{0.0103}^{0.01})$            &$-6.9\pm_{4.}^{3.9}\,(\pm_{6.3}^{6.5})$                          \\
$z_{\rm c}$                                                      &---                                                              &$1.86\pm_{0.16}^{0.11}\,(\pm_{0.27}^{0.3})$                      &$2.09\pm_{0.26}^{0.14}\,(\pm_{0.39}^{0.49})$                     &$4.\pm_{0.96}^{1.12}\,(\pm_{2.35}^{1.97})$                       \\
$M_B$ [mag]                                                              &$-19.356\pm_{0.011}^{0.01}\,(\pm_{0.021}^{0.021})$               &$-19.279\pm_{0.018}^{0.017}\,(\pm_{0.033}^{0.034})$              &$-19.256\pm_{0.018}^{0.018}\,(\pm_{0.035}^{0.037})$              &$-19.26\pm_{0.019}^{0.018}\,(\pm_{0.037}^{0.037})$               \\
$H_0$ [km/s/Mpc]                                                           &$70.45\pm_{0.4}^{0.38}\,(\pm_{0.76}^{0.77})$                     &$73.26\pm_{0.68}^{0.62}\,(\pm_{1.24}^{1.31})$                    &$74.07\pm_{0.68}^{0.68}\,(\pm_{1.31}^{1.43})$                    &$73.93\pm_{0.69}^{0.69}\,(\pm_{1.39}^{1.42})$                    \\
$\sigma_8$                                                       &$0.8102\pm_{0.0062}^{0.0055}\,(\pm_{0.0111}^{0.0119})$           &$0.8286\pm_{0.0068}^{0.0069}\,(\pm_{0.0138}^{0.0139})$           &$0.8344\pm_{0.0072}^{0.0069}\,(\pm_{0.014}^{0.0137})$            &$0.8326\pm_{0.0065}^{0.0066}\,(\pm_{0.0133}^{0.0131})$           

\\\bottomrule
\end{tabular}
\end{table}
\end{landscape}

\section*{Acknowledgements}
The work of L.A.A. is supported by the U.S. National Science
Foundation (NSF), Grant PHY-2412679. S.W. is supported by the AstroCom NYC program through NSF Grant AST-2219090.

\bibliography{references}

\end{document}